\documentclass[english]{spie}
\usepackage{mathptmx}

\usepackage[T1]{fontenc}
\usepackage[latin9]{inputenc}
\usepackage{graphicx}

\makeatletter
\newcommand{\lyxaddress}[1]{
\par {\raggedright #1
\vspace{1.4em}
\noindent\par}
}

\usepackage{url}

\@ifundefined{showcaptionsetup}{}{%
 \PassOptionsToPackage{caption=false}{subfig}}
\usepackage{subfig}
\makeatother

\usepackage{babel}
\begin{document}

\title{A simple technique for steganography}

\author{Adity Sharma, Anoo Agarwal and Vinay Kumar }

\maketitle

\lyxaddress{\begin{center}
Department of Electronics and Communication Engineering, \\Jaypee
University of Information Technology,Waknaghat,(H.P.)
\par\end{center}}

\section*{Abstract }

A new technique for data hiding in digital image is proposed in this
paper. Steganography is a well known technique for hiding data in an
image, but generally the format of image plays a pivotal role in it,
and the scheme is format dependent. In this paper we will discuss
a new technique where irrespective of the format of image, we can
easily hide a large amount of data without deteriorating the quality
of the image. The data to be hidden is enciphered with the help of
a secret key. This enciphered data is then embedded at the end of
the image. The enciphered data bits are extracted and then deciphered
with the help of same key used for encryption. Simulation results
show that Image Quality Measures of this proposed scheme are better
than the conventional LSB replacing technique. The proposed method
is simple and is easy to implement.

\section{Introduction}

Pertaining to requirement of privacy and security, there has been
a lot of progress in the field of text hiding and numerous new algorithms
are being developed enhancing the science of hiding information in
recent years. 

Two data hiding approaches are generally used. First, without considering
the image file format hide the information into the image file. In
this case the inherent file properties are not utilized and the approach
is more general in implementation. Second, designing the algorithm
or the approach based on the inherent image file format. In this case
least significant bit insertion, masking and filtering, algorithms
and transformations etcetera are utilized. In the present work we
use the first approach. 

Steganography \cite{Bender:1996:TDH:243519.243522,Cole2005,Johnson.1995,Jonathan2004,Sellars,Toshiyuki2004}
technique in technical field means having a piece of information and
hiding it into another image. It is similar to cryptography. Though
cryptography has its foundations in core mathematics and the encrypted
message (through cryptography), in case intercepted, can not be understood
unless the hiding mathematical tools are not applied in the correct
way. While steganography, in an essence, \textquotedbl{}camouflages\textquotedbl{}
a message to hide its existence and make it seem \textquotedbl{}invisible\textquotedbl{}
thus concealing the fact that a message is being sent altogether \cite{Johnson.1995}.
The advantage of this technique lies in the fact that though there
is a message but no one will (possibly) suspect it. Examples of this
technique are HP and Xeros printers, intelligence services and may
be in love letters. 

In the next section we present the proposed algorithm for data hiding.
Section 3 involves the experimental results in comparison with other
data hiding approaches. And finally concluding remarks are given in
Section 4.

\section{Proposed Algorithm }

The proposed algorithm is based on appending the data at the end of
the image file using file handling. The programming language used
is C and is chosen due to ease of understanding and portability. Data
hiding is used in conjunction with cryptography so that the information
is doubly protected, first it is encrypted and then hidden so that
an adversary has to first find the information and then has to decrypt
it. Attempt has been made to develop a technique which is independent
of any file format and can hide large amount of data with negligible
distortion.

\subsection{Hiding Principle }

The data to be hidden does not take advantage of the fact that there
are some areas present in the image which are not of too much importance,
it just appends the data in encrypted form at the end of the image.
It is a menu driven program which asks the user about the function
to be performed. If user wants to hide the data he\textbackslash{}she
has to provide the image file into which data is hidden, data file
whose data is to be hidden, the secret key and the marker. The image
and the data file should be kept in the Bin of the compiler for the
direct access. If they are not present in the bin a message invalid
will be flashed on the screen.Two file pointers, one pointing to the
image file and other pointing to the data file are initialized. The
image file is read until the EOF (end of file) character is encountered,
marker a special string is added at the end of the file which will
mark the beginning of the hidden text. 

The text to be hidden is first encrypted using a secret key that is
shared between the sender and the receiver. The encryption used is
monoalphabetic substitution i.e. if the character to be hidden is
\textquoteleft{}a\textquoteright{} and key is 6 it will be stored
as \textquoteleft{}g\textquoteright{}. Since most of the data to hidden
will be day to day English and if we append it directly at the end
of the image we can view the data contents by opening it in a Notepad
or WordPad. To overcome this problem value of key should be greater
than equal to 26 so that the data to be hidden moves out of the range
of the alphabets and could not be detected if opened in a Notepad
or WordPad. \textquoteleft{}A\textquoteright{} is the first alphabet
and needs key equal to 26 to move out of the alphabet range. This
is the maximum value required by any alphabet and i.e. why key is
taken to be greater or equal to 26.

\subsection{Retrieving Principle }

While retrieving the data receiver has to provide the image file,
the new data file into which the data retrieved is written, the secret
key and the marker. For the reliable recovery of the data from the
image the correct knowledge of the key and marker is must. 

Initial steps followed are same as those of the hiding process, two
file pointers, one pointing to the image file and other pointing to
the data file are initialized. The image file is read until the marker
is encountered. This is done using nested if loop, which keeps on
scanning the contents of the image file until the marker or EOF character
is encountered. The presence of marker indicates that some data is
present in the image. This indication is given by an indicator whose
value is set 0 initially. On encountering the marker it\textquoteright{}s
status is modified to 1 and the loop is exited. If the marker is not
found the value of the indicator remains 0. On coming out of the loop
the value of the indicator is checked, if it is 1 we apply decryption
algorithm on the data otherwise we exit the program and flash the
message that there is no data present in the image. The data after
the marker is decrypted using the priory knowledge of the key and
is written into a new file that could either be a .C file, .txt file
or any format. This new file created will be present in the BIN of
the compiler and can be accessed from there.

\subsection{Importance of the marker }

The marker added to the image file before adding the data has a two
way advantage first the data cannot be retrieved without it\textquoteright{}s
knowledge and second a different information can be hidden in the
same image by just changing the marker which is not possible with
other conventional techniques because in them the previous information
has to be overwritten. For e.g. in LSB substitution technique the
least significant bit is masked and the data to be hidden is placed
in it\textquoteright{}s position. Now if the same image is to be used
for hiding the data the information hidden in those least significant
bits will be overwritten by new information bits. Using this property
we can hide more than one data file at a time in the image out of
which only one is of importance. These different data files should
have different markers and different keys. While extracting the data
we can use the marker used for hiding the first data file but the
key used will be same as used for encrypting the data to be hidden.
The redundant data will make no sense on decryption and could be removed
leaving behind the required data. 

The secret key used for encryption can be obtained by studying the
frequency of occurrence of the different characters in the encrypted
data. However when different keys will be used it will be difficult
to obtain the key used for encrypting the data to be hidden.

\subsection{Implementation }

\begin{figure}
\centering

\includegraphics[width=12cm]{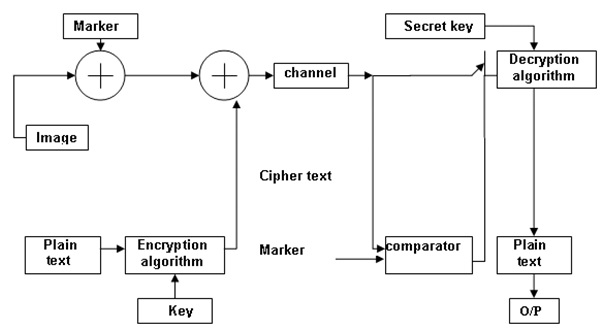}

\caption{\label{fig:Block-Diagram}Block Diagram}

\end{figure}

The algorithm discussed above can be realized with the help of electronic
circuitry (see Figure \ref{fig:Block-Diagram}). The text to be hidden
(plain text) is encrypted with the help of a secret key and converted
to cipher text. Marker, a special string is added at the end of the
image in which data is hidden. Cipher text is added to the image after
the marker and is then transmitted over a communication channel. 

At the receiver side, we have a comparator with the image contents
and marker as it\textquoteright{}s input. Initially the switch is
closed, the output of the comparator remains low until both the inputs
are not same. As soon as same inputs occur the output turns high and
the switch is closed. Image contents move through the switch and decryption
algorithm is applied to it which requires the knowledge of the secret
key. The text thus obtained is the hidden message in the image. Since
the marker is some characters long and according to the diagram only
one character is being compared at a time we need a storage device
for which we can use flip-flops; for example, if marker is 4 characters
long we will use 4 flip-flops and so on. These flip flops will be
required for both the inputs of the comparator.

\section{Experimental results }

For demonstrating the effectiveness of this algorithm to embed high
volume of data, we have used several images with different file formats.
The result is analyzed with other hiding approaches and superior performance
in our technique is quite obvious as it supports all image file formats
and huge volume of data can be embedded into image without distorting
it.

Hidden data is well protected when key is greater than 26. Moreover
using this algorithm same image can be used to hide more than one
data file at a time once which is not possible with many existing
data hiding techniques . Although the file size increases after the
appending of the data but it is not important if the amount of data
that can be hidden is considered, moreover the adversary could not
make it out unless he has the access to the original image without
data. 

\begin{figure}
\centering

\subfloat[Without encrypted data]{\includegraphics[width=6cm]{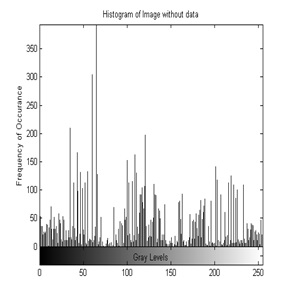}

} \subfloat[With encrypted data.]{\includegraphics[width=6cm]{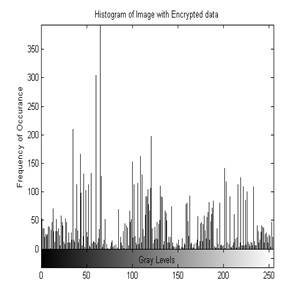}

}

\caption{\label{fig:Histogram-plots-with}Histogram plots with data and without
encrypted data}

\end{figure}

\begin{figure}
\centering

\includegraphics[width=6cm]{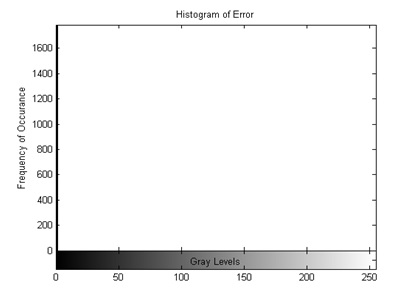}

\caption{\label{fig:Histogram-of-error}Histogram of error}

\end{figure}

The histogram plots of the images showing the gray levels and their
frequency of occurrence are plotted and found to be identical for
the image with data and that without data (See Figures \ref{fig:Histogram-plots-with}
and \ref{fig:Histogram-of-error}), the histogram of error shows that
there is no difference between the two.

\section{Conclusion }

In this method, a scheme of embedding large volume of data is presented.
The scheme has following features: supports almost all image file
formats ,relatively large volume of data can be embedded in comparison
to other data hiding techniques and image distortion is negligible,
the data hiding ratio is improved considerably as the image pixel
values are not changed. Use of a marker and encryption provides an
extra layer of security. Future work is concerned with improving the
security of the scheme by using advanced cryptographic techniques.

\end{document}